\documentclass[%
 reprint,
 amsmath,amssymb,
]{revtex4-2}

\usepackage{graphicx}%
\usepackage{dcolumn}%
\usepackage{bm}%
\usepackage{hyperref}%

\usepackage{xcolor}

\newcommand{\vect}{\boldsymbol}
\newcommand{\mobD}{\Lambda_\mathrm{d}}
\newcommand{\mobR}{\Lambda_\mathrm{r}}
\newcommand{\mobRA}{\mobR^\mathrm{a}}
\newcommand{\mobRP}{\mobR^\mathrm{p}}
\newcommand{\etaR}{\eta_\mathrm{r}}
\newcommand{\mean}[1]{\langle #1 \rangle}
\newcommand{\figref}[1]{Fig.~\ref{#1}}
\newcommand{\Eqref}[1]{Eq.~\eqref{#1}}
\newcommand{\cIn}{c_\mathrm{in}}
\newcommand{\cOut}{c_\mathrm{out}}
\newcommand{\cSpin}{c_\mathrm{sp}}
\newcommand{\Rcrit}{R_\mathrm{crit}}
\newcommand{\kCrit}{k_\mathrm{max}}
\newcommand{\RcritPas}{\Rcrit^\mathrm{pas}}
\newcommand{\kBT}{k_\mathrm{B} T}

\begin{document}
\preprint{APS/123-QED}

\title{Nucleation of chemically active droplets}%

\author{Noah Ziethen}
 \author{Jan Kirschbaum}
\author{David Zwicker}%
 \email{david.zwicker@ds.mpg.de}
\affiliation{%
 Max Planck Institute for Dynamics and Self-Organization\\
 Am Faßberg 17 37077 Göttingen
}%

\date{\today}%

\begin{abstract}
Driven chemical reactions can control the macroscopic properties of droplets, like their size. Such active droplets are critical in structuring the interior of biological cells. Cells also need to control where and when droplets appear, so they need to control droplet nucleation. Our numerical simulations demonstrate that reactions generally suppress nucleation if they stabilize the homogeneous state. An equilibrium surrogate model reveals that reactions increase the effective energy barrier of nucleation, enabling quantitative predictions of the increased nucleation times. Moreover, the surrogate model allows us to construct a phase diagram, which summarizes how reactions affect the stability of the homogeneous phase and the droplet state. This simple picture provides accurate predictions of how driven reactions delay nucleation, which is relevant for understanding droplets in biological cells and chemical engineering. 

\end{abstract}

\maketitle

\tableofcontents

\section{Introduction}

Droplets forming by phase separation are crucial to spatially structure the interior of biological cells, e.g., to separate molecules, control reactions, and exert forces~\cite{Brangwynne_Eckmann_Courson_Rybarska_Hoege_Gharakhani_Julicher_Hyman_2009,Hyman2014,Banani_Lee_Hyman_Rosen_2017,Dignon2020,Su2021}.
Cells control phase separation using actively driven chemical reactions, which often include enzymes that modify biomolecules involved in phase separation~\cite{Hondele2019,Snead2019,Snead2019,Soeding_Zwicker_Sohrabi-Jahromi_Boehning_Kirschbaum_2019}.
Theoretical work showed that such reactions can control droplet sizes and their general macroscopic behavior~\cite{Zwicker_Decker_Jaensch_Hyman_Juelicher_2014,Zwicker_Hyman_Juelicher_2015,Zwicker_Seyboldt_Weber_Hyman_Juelicher_2017,Wurtz_Lee_2018_2,Weber_Zwicker_2019,Kirschbaum_Zwicker_2021,Zwicker2022a}.
In contrast, how these droplets emerge is little understood.
Experiments suggest that droplets form by nucleation~\cite{Shimobayashi2021}, but to what extent reactions can regulate nucleation is unclear.

Nucleation is a stochastic process that relies on thermal fluctuations to create a sufficiently large nucleus that can grow spontaneously~\cite{Sear2007a, Xu2014, Kalikmanov2013}.
This is because creating the droplet interface costs energy, so tiny droplets generally dissolve.
Classical nucleation theory predicts that the typical nucleation time scales exponentially with the energy barrier associated with the critical nucleus.
While this theory is well-understood for passive systems, it is unclear how it can be extended to active systems, where free energies are generally unavailable.
To overcome this challenge, we use an equilibrium surrogate model to reveal how driven reactions controlling droplet size suppress nucleation substantially.

\section{Model}
\label{sec:model}
We study an isothermal fluid comprised of precursor material $A$ that can convert into droplet material $B$ by chemical reactions.
For simplicity, we consider an incompressible fluid where both species have equal molecular volume~$\nu$, so the state of the system is characterized by the concentration~$c(\vect r, t)$ of species $B$, while the concentration of $A$ is $\nu^{-1} - c(\vect r, t)$.
The dynamics are governed by the continuity equation
\begin{align}
    \partial_t c + \nabla \cdot \vect{j} = s \; ,
    \label{eq:cahn-hilliard-source}
\end{align}
where $\vect j$ denotes the diffusive exchange flux and the source term~$s$ describes chemical transitions.

The passive diffusive flux~$\vect j$ is driven by the gradient of the chemical potential, $\vect j = - \mobD \nabla \mu + \vect\eta$, where $\mobD$ is the diffusive mobility and $\vect\eta$ is the diffusive thermal noise, which obeys $\mean{\eta_i(\vect r, t)}=0$ and the fluctuation dissipation theorem $\mean{\eta_i(\vect r, t) \eta_j (\vect r', t')} = 2 k_{\mathrm{B}} T \mobD \delta_{ij} \delta\left(\vect r - \vect r'\right) \delta\left(t-t'\right)$~\cite{nonEqThermo}.
The exchange chemical potential, $\mu = \delta F[c]/\delta c$, follows from the free energy functional
\begin{align}
    F[c] = \int \left[ f(c)+\frac{\kappa}{2}|\nabla c|^2 \right] \mathrm{d} \vect{r} \;,
\end{align}
where $f(c)$ is the local free energy density and $\kappa$ penalizes compositional gradients.
For simplicity, we consider
\begin{align}
	f(c) = 
		a_1 c
		-\frac{a_2}{2}\left(c - \frac{1}{2\nu}\right)^2 
		+ \frac{a_4}{4}\left(c - \frac{1}{2\nu}\right)^4 
	\label{eq:f_loc}
	\;,
\end{align}
where $a_1$, $a_2$, and $a_4>0$ are phenomenological coefficients.
Without reactions ($s=0$), Eqs.~\eqref{eq:cahn-hilliard-source}--\eqref{eq:f_loc} describe passive phase separation with a critical point at $c_\mathrm{crit}=\frac{1}{2\nu}$ for $a_2=0$.
For $a_2>0$, the spinodal line is given by $c_\textrm{sp}=\frac{1}{2\nu} \pm \sqrt{a_2/ (3 a_4)}$ and the binodal is defined by coexisting equilibrium concentrations $\cOut=\frac{1}{2\nu}-\sqrt{a_2/a_4}$ and $\cIn=\frac{1}{2\nu}+\sqrt{a_2/a_4}$ in dilute and dense phases, respectively.

\begin{figure}
    \centering
    \includegraphics[width=\columnwidth]{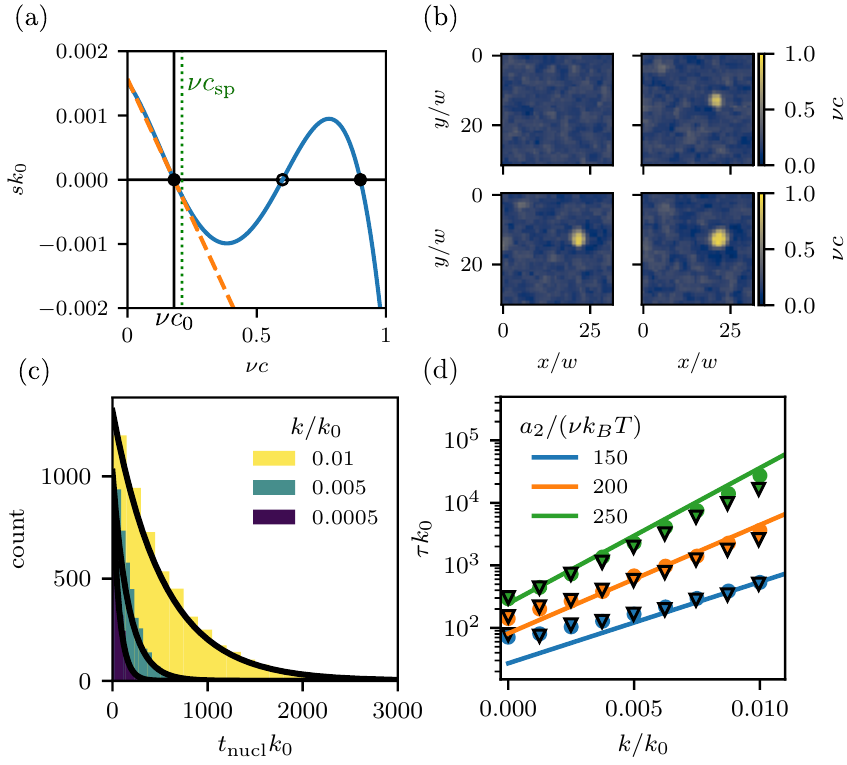}
    \caption{
    \textbf{Chemical reactions increase nucleation times.}
    (a) Reaction flux~$s$ as a function of the concentration $c$ of a homogeneous system for the full (solid blue line, \Eqref{eq:react_cons}) and linearized model (dashed orange line).
    The spinodal concentration~$\cSpin$ of the passive system (dotted green line), the two stable fixed points (filled disks), and the unstable fixed point (open circle) are marked.
    (b) Snapshots of the concentration field~$c$ of droplet material obtained from stochastic numerical simulation in two dimensions. 
    The time between snapshots is $10 /k_0$ and the interaction strength is $a_2=200\,\nu \kBT$. 
    (c) Distribution of measured nucleation times~$t_\mathrm{nucl}$ in the linearized model for various reaction rates $k$ for $a_2 = 150 \, \nu \kBT$.
    Black lines show exponential distributions of equivalent mean $\tau=\mean{t_\mathrm{nucl}}$.     	(d) Nucleation time $\tau$ as a function of $k$ for the full model (disks, $k=-s'(c_0)\propto\mobRA$) and the linearized reactions (triangles) for various interaction strengths~$a_2/(\nu k_BT)$.
    Solid lines show predictions of \Eqref{eq:arrhenius} with $A$ as a single fit parameter for all curves.
    (a--d) Additional parameters are $a_1 \nu=-1.34\,a_2$, $a_4=4\,a_2 \nu$, $\mobRP/\mobRA=0.0311$, $\Delta\mu \nu=1.46\,a_2$, $\nu=w^2$,
    $w=\sqrt{2\kappa/a_2}$, and $k_0=\mobD a_2 w^{-2}$.
    }
    \label{fig:numerics}
\end{figure}

The system becomes active when phase separation is augmented by driven chemical reactions.
We first consider a reaction flux~$s$ comprising passive conversion of $A$ and $B$ as well as an active conversion involving chemical energy~$\Delta \mu$ provided by a fuel~\cite{Kirschbaum_Zwicker_2021},
\begin{align}
    s(c) &= -\mobRP\mu - \mobRA c \cdot \bigl(\mu + \Delta\mu\bigr)  + \etaR(c)
    \;,
    \label{eq:react_cons}
\end{align}
where $\mobR^\mathrm{p}$ and $\mobR^\mathrm{a}$ determine the rates of the respective reactions and $\etaR$ models thermal fluctuations.
Motivated by enzymes that co-localize with the droplet, we scale the rate of the active reaction with the concentration~$c$ of the droplet material.
This choice allows stationary states where droplet material~$B$ turns into precursor~$A$ inside droplets, while $B$ is replenished outside, thus controlling droplet size~\cite{Kirschbaum_Zwicker_2021}.
The reactive thermal noise $\etaR$ obeys $\mean{\etaR(\vect r, t)}=0$ and $\mean{\etaR(\vect r, t) \etaR (\vect r', t')} = 2 k_{\mathrm{B}} T \mobR \delta\left(\vect r - \vect r'\right) \delta\left(t-t'\right)$ with $\mobR(c)=\mobRP + \mobRA c $.
Active droplets are only stable if reactive fluxes are weak compared to diffusive fluxes~\cite{Kirschbaum_Zwicker_2021}.
Consequently, the reactive noise~$\etaR$ is much weaker than the diffusive noise~$\vect\eta$ and we neglect it in the following.
\figref{fig:numerics}a shows that the reaction flux~$s$ given in \Eqref{eq:react_cons} is a non-monotonous function of the composition~$c$.
In particular, there are two stable homogeneous stationary state, which correspond to (meta-)stable dilute and dense systems.
The main question in this paper concerns how active droplets nucleate from the dilute homogeneous state $c(\vect r) = c_0$.

\section{Results}
\subsection{Chemical reactions hinder nucleation}
\label{sec:num-sim}
To investigate nucleation, we first perform numerical simulations of Eqs.~\eqref{eq:cahn-hilliard-source}--\eqref{eq:react_cons} in two-dimensional system with periodic boundary conditions~\cite{Zwicker_2020}; see \figref{fig:numerics}b.
Repeating the simulations many times, we observe that the first droplet nucleates at random times~$t_\mathrm{nucl}$; see \figref{fig:numerics}c.
Assuming an exponential distribution of $t_\mathrm{nucl}$, we define the nucleation time~$\tau$ as the ensemble average of $t_\mathrm{nucl}$.
\figref{fig:numerics}d shows that $\tau$ increases for stronger interactions (larger $a_2$), as expected for nucleation of passive droplets~\cite{Arrhenius}.
More importantly, larger reaction rates~$\mobRA$ lead to longer nucleation times~$\tau$, indicating that active chemical reactions hinder nucleation.
This result can be understood intuitively since the reactions stabilize the homogeneous state; see \figref{fig:numerics}a.
They thus help to dissolve a small accumulation of droplet material~$B$, reducing the probability that a critical nucleus forms.

\subsection{Surrogate equilibrium system reveals additional energy barrier} 
\label{sec:mapp}

To quantitatively understand the effect of driven reactions on nucleation, we next map our system to an approximate equilibrium system.
To do this, we linearize the reaction flux~$s$ around the dilute homogeneous stationary state $c_0$,
\begin{align}
	s_\mathrm{lin}(c) &= k (c_0 - c)
& \text{with} &&
	k&=-s'(c_0)
\end{align}
where $k>0$ for a stable state; see \figref{fig:numerics}a.
\figref{fig:numerics}d shows that the linearized reactions influence the nucleation time~$\tau$ similarly to the full reaction flux~$s$.
The linearization allows us to map the dynamics given by \Eqref{eq:cahn-hilliard-source} to a passive system, $\partial_t c \approx \mobD \nabla^2 \delta \tilde F[c]/\delta c$, with the augmented free energy functional
\begin{align}
 	\tilde F[c] %
 	= F[c]  + \frac{k}{2 \mobD} \int \bigl[c(\vect r) - c_0\bigr]\Psi(\vect r) \, \mathrm{d}\vect{r}
	\label{eq:tot_en_func}
	\;,
\end{align}
where $\Psi$ obeys the Poisson equation $\nabla^2 \Psi = c_0 - c(\vect r)$ and thus mediates long-ranged, Coulomb-like interactions~\cite{Liu_Goldenfeld_1989, Christensen_Elder_Fogedby_1996, Muratov_2002}.

\begin{figure}
    \centering
    \includegraphics[width=\columnwidth]{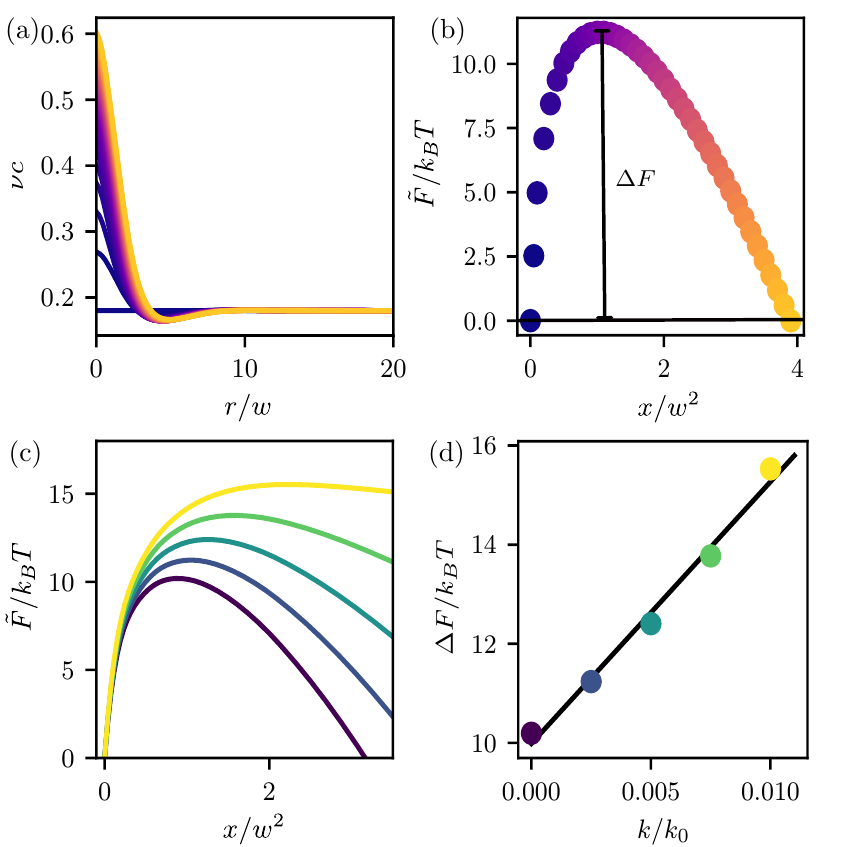}
    \caption{\textbf{Reactions increase free energy barrier~$\Delta F$ of surrogate equilibrium model.}
    (a) Radial concentration profiles $c(r)$ minimizing the free energy~$\tilde F$ given by \Eqref{eq:tot_en_func} at various fixed values of reaction coordinate~$x$ (colors correspond to panel b).
    (b) $\tilde F$ as a function of $x$ with $\Delta F$ indicated.
    (c) $\tilde F(x)$ for various reaction rates~$k$ (colors correspond to panel d).
    (d) $\Delta F$ as a function of $k$.
    (a--d) Model parameters are $c\nu =0.18$, $L=100\,w$, $a_2=250\,\nu \kBT$, $k=0.0025\,k_0$ (for panels a and b) and given in \figref{fig:numerics}.   %
    }
    \label{fig:optimized-profiles}
\end{figure}

We use the equilibrium surrogate model to investigate the energy landscape of nucleation.
In particular, we use \Eqref{eq:tot_en_func} to map the minimal energy path connecting the metastable homogeneous state with the equilibrium state containing one droplet using a proxy for droplet size as a reaction coordinate~$x$; see Appendix~I.
For each value of $x$, we use a constrained optimization to determine the spherically symmetric composition~$c(r)$ that minimizes the energy~$\tilde F$ given by \Eqref{eq:tot_en_func}.
\figref{fig:optimized-profiles}a shows that the resulting profiles feature an increasing density peak, analogous to passive systems~\cite{Kalikmanov2013}.
However, the nucleus is also surrounded by a depletion zone originating from chemical reactions.
The sequence of profiles defines the minimal energy path, from which we obtain the energy barrier~$\Delta F$ as the difference between the maximal energy and the energy~$\tilde F(x=0)$ of the homogeneous state; see \figref{fig:optimized-profiles}b.
\figref{fig:optimized-profiles}c shows that the energy barrier~$\Delta F$ depends on the reaction rate~$k$ and this dependence is approximately linear (\figref{fig:optimized-profiles}d), suggesting that the long-ranged term in \Eqref{eq:tot_en_func} could explain the suppressed nucleation caused by reactions.

We hypothesize that the increasing energy barriers explain how larger reaction rates~$k$ lead to longer nucleation times~$\tau$ (see \figref{fig:numerics}d).
Nucleation theory predicts that $\tau$ increases exponentially with the energy barrier~$\Delta F$~\cite{Arrhenius},
\begin{align}
    \tau = A \exp{\left(\frac{\Delta F}{k_{\rm B} T}\right)}
    \label{eq:arrhenius}
    \;,
\end{align}
where $A$ is a kinetic prefactor.
\figref{fig:numerics}d shows that this relation explains the numerical data, particularly for larger~$\Delta F$ at higher $k$ and $a_2$.
The deviation at smaller $k$ are expected since the assumptions leading to \Eqref{eq:arrhenius} break down for smaller $\Delta F$~\cite{Kalikmanov2013}.
We conclude that the energy barriers derived from the equilibrium surrogate model explain how nucleation times increase with reaction rates.

\subsection{Classical nucleation theory leads to phase diagram extended by chemical reactions}
\label{sec:cnt}
\begin{figure} 
    \centering
    \includegraphics[width=\columnwidth]{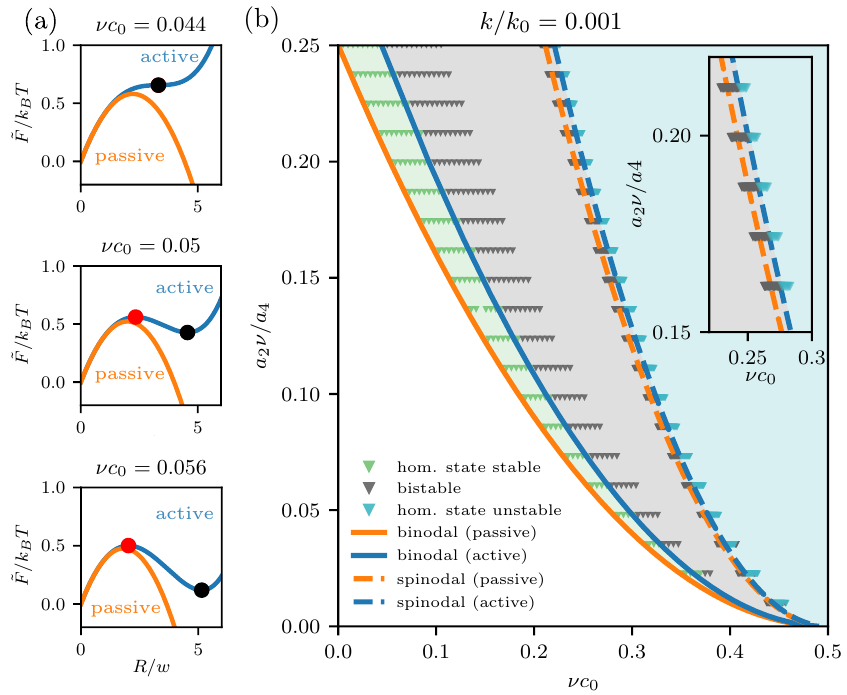}
    \caption{\textbf{Extended phase diagram accounting for reactions.}
    (a)~Free energy~$\tilde F$ of the surrogate equilibrium model approximated by \Eqref{eqn:energy_nucleus} as a function of the nucleus radius~$R$ for decreasing concentrations $c_0$ of the homogeneous state (bottom to top) and $a_2 \nu/a_4=0.25$.
    (b) Extended phase diagram indicating the stability of the homogeneous and droplet state as a function of $c_0$ and $a_2 \nu/a_4$ in a passive (orange lines, $k=0$) and active system (blue lines, $k=10^{-3}k_0$).
    The droplet state is (meta-)stable right of the solid (binodal) lines, while the homogeneous state is stable left of the dashed (spinodal) lines.
    Behavior of numerical simulations (symbols) for $k=10^{-3}k_0$ corroborate the results.
    Inset shows zoomed in region around the spinodal.
    (a--b) Additional model parameters are given in \figref{fig:numerics}.
    }
    \label{fig:phase_diagram}
\end{figure}

Motivated by the success of nucleation theory, we next approximate the minimal free energy path using the radius~$R$ of a droplet as a reaction coordinate.
Assuming that the droplet with homogoneous concentration~$\cIn$ is embedded in an infinite system of concentration~$c_0$, the free energy~$\tilde F$ can be separated into contributions of bulk phases, interface, and chemical reactions,
\begin{align}
  \label{eqn:energy_nucleus}
   \tilde F(R) \approx -g V + \gamma A  + F_\mathrm{react}(R)
   \;,
\end{align}
where $V= \pi R^2$ and $A=2\pi R$ in two dimensions.
Classical nucleation theory implies the free energy difference $g=f(c_0) - f(\cIn) + \mu(c_0)(\cIn - c_0)$ between the phases and surface tension $\gamma=\sqrt{8 \kappa c_2^3}/(3c_4)$~\cite{Weber_Zwicker_2019}, which is a good approximation for $c_0 \approx \cOut$.
We show in Appendix~II that the free energy associated to reactions is approximately
\begin{align}
	F_\mathrm{react}(R)\approx \frac{\pi (\cIn - c_0)^{2}}{16\mobD} k R^{4}
	\;,
\end{align}
where we neglected terms proportional to $k(\cOut - c_0)^2$.
Without reactions ($k=0$), $\tilde F(R)$ given by \Eqref{eqn:energy_nucleus} has a single maximum at the critical radius $\RcritPas = \gamma/g$ with a corresponding energy barrier~$\Delta F = \pi\gamma^2/g$; see \figref{fig:phase_diagram}a.
Once nuclei exceed this critical size (by nucleation), they grow indefinitely.
In contrast, \Eqref{eqn:energy_nucleus} predicts that reactions ($k>0$) increase the free energy of large droplets, implying a minimum at finite radius~$R_\mathrm{stab}\approx (c_0-\cIn)^{-2}[8g\mobD/k]^{1/2}$ corresponding to stable droplets~\cite{Zwicker_Hyman_Juelicher_2015}.
Concomitantly, the energy barrier $\Delta F$ is elevated, consistent with increased nucleation times.
Approximating the barrier by $\tilde F(\RcritPas)$, we find
\begin{equation}
	\Delta F \approx  \frac{\pi \gamma^2}{g} \left[
		1
		+ k  \frac{\gamma^2 (\cIn - c_0)^{2}}{16 g^3 \mobD}
		\right]
	\;,
\end{equation}
which explains the linear dependence of $\Delta F$ on $k$ observed in \figref{fig:optimized-profiles}d. 
Taken together, this simplified picture demonstrates how large rates~$k$ gradually disfavor the droplet state until only the homogeneous state remains stable at $k > \kCrit$ with
$\kCrit = 32 \mobD  g^3/ [27 \gamma ^2 (\cIn - c_0)^2]$.

Finally, we use the simplified free energy of the equilibrium surrogate model to study the influence of the concentration~$c_0$ of the homogeneous state.
In particular, we determine the minimal value of $c_0$ beyond which the droplet state can be stable as a function of the interaction parameter~$a_2$.
In passive systems ($k=0$), the resulting line corresponds to the binodal curve.
In active system ($k>0$), this line is shifted to larger concentrations, thus enlarging the region where the homogeneous system is stable.
The homogeneous system becomes unstable at the spinodal line, which can be determined from a linear stability analysis of \Eqref{eq:cahn-hilliard-source}; see Appendix~III.
\figref{fig:phase_diagram}b shows that these predictions based on \Eqref{eqn:energy_nucleus} agree with numerical simulations probing the stability of the homogeneous and droplet state.
Both predictions illustrate how driven chemical reactions destabilize the droplet state.

\section{Discussion}

In summary, we illuminated how driven chemical reactions affect the phase diagram of phase separating systems.
To do this, we exploited the equilibrium surrogate of the active system to show how reactions favor the homogeneous state relative to the droplet state, which explains the suppressed nucleation qualitatively.
Similar behavior was found for equilibrium system with true long-ranged interactions~\cite{Muratov_2002}.
Although the modified phase diagram was derived from the surrogate model, it is not a thermodynamic phase diagram of the phase separating system with driven reactions.
For instance, the compositions of the coexisting phases at the interface are still governed by the binodal and tie lines of the passive phase diagram~\cite{Weber_Zwicker_2019}.
The energy barrier associated with reactions depends linearly on their rate $k$, likely because reactions are weak and the system is dominated by phase separation.
This implies that $k$ decreases nucleation rates exponentially.
We showed that this dependence persists for thermodynamically consistent reactions and expect that it is a general feature of phase separating systems with reactions that have a stable dilute phase.
Moreover, since our derivation of the influence of the reactions is independent of the details of the free energy density, we expect that reactions suppress nucleation in a wide range of phase separating systems.

We presented results for the simple case of a binary fluid in two dimensions.
While we expect that active reactions also slow nucleation in more complicated situations, it will be vital to extend our theory to three dimensions (e.g., to capture spontaneous divisions~\cite{Zwicker_Seyboldt_Weber_Hyman_Juelicher_2017}) and many components (allowing for additional stable stationary states~\cite{Kirschbaum_Zwicker_2021,Zwicker2022a}).
For better quantitative agreement, it might also be necessary to improve our treatment of nucleation theory, e.g., by describing how reactions affect the curvature of the surrogate free energy, which affects nucleation rates via the Zeldovich factor~\cite{zeldovich1942theory}.
However, the ultimate test of our theory will come from experiments, either from existing active droplets in biological cells~\cite{Wurtz_Lee_2018} or in promising synthetic systems~\cite{Donau2022,Nakashima2021}.
Experiments in cells also suggest that more complex behaviors are possible, including periodic nucleation~\cite{Yan2022a} and multi-step nucleation for fiber formation~\cite{Martin2021}, which might involve secondary nucleation~\cite{Weber2019a}.
In these situations, heterogeneous nucleation is likely relevant~\cite{Qi2021, Wei_Chang_Shimobayashi_Shin_Strom_Brangwynne_2020} and there are examples where nucleation is controlled by catalytically-active nucleation sites~\cite{Zwicker_Decker_Jaensch_Hyman_Juelicher_2014,Zwicker_Baumgart_Redemann_Mueller-Reichert_Hyman_Juelicher_2018}.
Taken together, our approach of an equilibrium surrogate model will likely prove vital for studying nucleation in these more challenging situations.

\begin{acknowledgments}
We thank Kristian Blom and Aljaz Godec for helpful discussions and Gerrit Wellecke for a critical reading of the manuscript.
We gratefully acknowledge funding from the Max Planck Society and the German Research Foundation (DFG) under grant agreement ZW 222/3.
\end{acknowledgments}

\begin{appendix}

\section{Constrained optimizations to uncover minimal energy path}
The minimal energy path comprises a sequence concentration profiles that connects the homogeneous state to a stationary droplet.
We determine the path using a measure for the mass concentrated in the nucleus, $x[c] = \int \frac{1}{2}[1 + \tanh(\beta(c-\frac{1}{\nu 2}))] \mathrm{d}V$ with $\beta=10$, as a reaction coordinate~$x$.
We determine the minimal energy path by minimizing the free energy $\tilde F$ (given in Eq. 6) with constrained values of~$x$.
We impose a value~$x_0$ of the reaction coordinate using a Lagrange multiplier~$\lambda$ and minimize
\begin{align}
    \hat F[c, \lambda] = \tilde{F}[c] - \lambda (x[c] - x_0)
\end{align}
by evolving corresponding partial differential equations
\begin{subequations}
\begin{align}
    \partial_t c &= \Lambda_D \nabla^2\frac{\delta \hat F}{\delta c}
\\
    \partial_t \lambda &= -\Lambda_L \frac{\delta \hat F}{\delta \lambda}
    \;,
\end{align}
\end{subequations}
which corresponds to conserved and non-conserved dynamics with mobilities $\Lambda_D$ and $\Lambda_L$, respectively.
We use $\Lambda_D=1$ and $\Lambda_L=100$, which proved a good compromise between speed and stability.
Using this procedure, we determine the radially symmetric profile~$c(r)$ that minimize $\hat F$ for each value $x_0$ of the constraint, which yields the minimal free energy path.
The profile with the largest energy~$\tilde F$ corresponds to the saddle point and thus the critical nucleus.

\section{Approximation of the non-local free energy}
We here approximate the non-local term in the free energy, which describes the effect of the reactions in the surrogate equilibrium model.
We consider a spherical droplet of radius $R$ and constant concentration $\cIn$ embedded in a large dilute phase with a spherically-symmetric concentration profile $\cOut(r)$, which we will determine approximately for small $\cOut - c_0$. 
The time evolution of the concentration field $c(r,t)$ is given by the Cahn-Hilliard equation with an additional source term taking into account the linearized reaction rates:
\begin{align}
    \partial_{t} c=\mobD \nabla^{2} \mu-k\left(c-c_{0}\right)
    \;,
    \label{eq:dyneq}
\end{align}
where  $\mu$ is the exchange chemical potential, $c_0$ the background concentration, and $k$ the linear reaction rate.
Since $c(r\rightarrow \infty) = c_0$, we linearize $\mu(c)$ around $c=\cOut$,
\begin{align}
    \mu(r) = \mu(c(r)) = \left.\frac{\partial \mu}{\partial c}\right|_{c=\cOut} \bigl(c(r)-\cOut\bigr)
    \;,
\end{align}
assuming $c_0 - \cOut$ is small. 
Inserting this in Eq. \ref{eq:dyneq}, we find the reaction-diffusion equation
\begin{align}
    \partial_t c = \Lambda \mu' \nabla^2c-k(c-c_0),
\end{align}
where $\mu'= \left.\frac{\partial \mu}{\partial c}\right|_{c=\cOut}$.
In the stationary state,
\begin{align}
    0 = \nabla^2 c - \xi^2(c-c_0),
\end{align}
where $\xi^2=k/(\Lambda \mu')$ is the inverse of a reaction-diffusion length scale.
Using the boundary conditions $c(R)=\cOut$ and $c(\infty)=c_0$, we find
\begin{align}
   c(r) = \frac{(\cOut-c_0) K_0(\xi  r)}{K_0(\xi  R)}+c_0
   \;,
\end{align}
where $K_0$ denotes a modified Bessel function of the second kind.
Note that the limit $\xi \rightarrow 0$ is not applicable since the passive steady state concentration profile is not compatible with the chosen boundary conditions.
In the numerical solution of the concentration profile it becomes apparent that the boundary concentration $\cOut$ depends on $\xi$ which is not captured by our approximation. We will be therefore limited to parameters $\xi>\epsilon$.  %
Taken together, the approximated concentration profile of the droplet reads
   \begin{align}
    c( r) = \begin{cases}
        \cIn & r\leq R  \\[7pt]
        \dfrac{(\cOut-c_0) K_0(\xi  r)}{K_0(\xi  R)}+c_0 & r>R \;.
                \end{cases}
   \label{eq:conc-profile}
   \end{align}
We next use this profile to evaluate the non-local part of the free energy density $\tilde F$ given by Eq. 6,
   \begin{align}
       F_\textrm{react}=  \frac{k}{2 \mobD} \int \bigl[c(\vect r) - c_0\bigr]\Psi(\vect r) \, \mathrm{d}\vect{r} \;,
    \end{align}
    where $\Psi$ obeys the Poisson equation $\nabla^2 \Psi = c_0 - c(\vect r)$.
Using the definition of $\Psi$, we find
    \begin{align}
       F_\textrm{react}=  -\frac{k}{2 \mobD} \int \Psi(\nabla^2 \Psi)  \, \mathrm{d}\vect{r} \;.
    \end{align}
    Integrating py parts and dropping the boundary term, we obtain
    \begin{align}
        F_\textrm{react}
                            &=\frac{k}{2 \mobD} \int (\nabla \Psi) (\nabla \Psi)  \, \mathrm{d}\vect{r} \;.
    \end{align}
For the radial-symmetric system in two dimensions that we consider, we find
    \begin{align}
       F_\textrm{react} = \frac{\pi k}{\mobD} \int\left(\partial_{r} \Psi\right)^{2} \, r\mathrm{d}r
       \label{eq:FNL}
   \end{align}
   with $\nabla^{2} \Psi=\partial_{r}^{2} \Psi+\frac{1}{r} \partial_{r} \Psi=-\left(c-c_{0}\right)$.
   Defining $\Omega=\partial_r \Psi$, we get
\begin{align}
    \partial_{r} \Omega+\frac{\Omega}{r}=-(c(r)-c_0).
    \label{eq:poissoneq}
\end{align}
Using Eq.~\eqref{eq:conc-profile}, we obtain
\begin{align}
    \Omega = \begin{cases}
        -\dfrac{(\cIn-c_0)r}{2} & r<R  \\[7pt]
        \dfrac{d_1}{r}-\dfrac{({c_0}-\cOut) K_1(\xi  r)}{\xi  K_0(\xi  R)} & r>R \;,
    \end{cases}
\end{align}
where $d_1$ is an integration constant.
Denoting the two branches by $\Omega_\mathrm{in}$ and $\Omega_\mathrm{out}$, we find from Eq. \ref{eq:FNL}
\begin{widetext}
\begin{align}
    F_\textrm{react} &= \frac{\pi k}{\mobD} \int_0^R \Omega_\textrm{in}^2 r \mathrm{d}r + \frac{\pi k}{\mobD} \int_R^{L} \Omega_\textrm{out}^2 r \mathrm{d}r 
\notag\\
           &=  \frac{\pi k}{\mobD} \frac{(\cIn-c_0)^2}{16}R^4
           + \frac{\pi k}{\mobD} \left(
           	\int_R^{L} \frac{d_1^2}{r} \mathrm{d}r 
			- \int_R^{L} d_1\frac{({c_0}-\cOut) K_1(\xi  r)}{\xi  K_0(\xi  R)}\mathrm{d}r 
			+ \int_R^L \left(\frac{({c_0}-\cOut) K_1(\xi  r)}{\xi  K_0(\xi  R)}\right)^2 r \mathrm{d}r\right)
           \;,
           \label{eqn:FreactFull}
\end{align}
\end{widetext}
where $L$ denotes the radius of the spherical system.
Since the integrals must converge in the limit $L \to \infty$, we conclude $d_1=0$.
Hence, the second term of Eq.~\eqref{eqn:FreactFull} reduces to
\begin{align}
	F_\textrm{react}^\textrm{out} &\approx \frac{\pi k (c_0 - \cOut)^2 R^2 }{\mobD \xi^2} 
	\left[
        \frac{1}{2}+\frac{K_1(\xi R)}{R\xi K_0(\xi R)}-\frac{K_1(\xi R){}^2}{2 K_0(\xi R){}^2}
	\right]
\end{align}
in the limit of large systems, $L \to \infty$.
For large droplet radii ($R\to\infty$), this contribution scales as $R$ and can thus be neglected compared to the contribution from inside the droplet, which scales as $R^4$.
Additionally, the prefactors are different, $({c_0}-\cIn)^2 \gg ({c_0}-\cOut)^2$.
Since critical radii become large close to the binodal concentration (where classical nucleation is valid), we can neglect the contribution from the outside to find Eq. 9.
\section{Spinodal line from linear stability analysis}
We perturb the dynamic equation given by Eq. 3 around the homogeneous state using the ansatz
\begin{align}
    c(\vect r, t) = c_0 + \epsilon e^{\omega t + i \vect q \vect r} \;.
\end{align}
To linear order in $\epsilon$, we find
\begin{align}
    \omega = \mobD a_{2} q^{2} - \frac{3 \mobD a_{4} \vect q^{2} \left(2 c_{0} \nu - 1\right)^{2}}{4 \nu^{2}} - \mobD \kappa \vect q^{4} - k.
\end{align}
The maximal growth rate $\omega_\mathrm{max}$ reads
\begin{align}
\nonumber
    \omega_\mathrm{max} &= \frac{\mobD c_{2} \left(4 a_{2} \nu^{2} - 12 a_{4} c_{0}^{2} \nu^{2} + 12 a_{4} c_{0} \nu - 3 a_{4}\right)}{8 \kappa \nu^{2}} 
\\ \nonumber  &\quad
    - \frac{3 \mobD c_{4} \phi_{0}^{2} \left(4 a_{2} \nu^{2} - 12 a_{4} c_{0}^{2} \nu^{2} + 12 a_{4} c_{0} \nu - 3 a_{4}\right)}{8 \kappa \nu^{2}} \\\nonumber
    &\quad+ \frac{3 \mobD c_{4} \phi_{0} \left(4 a_{2} \nu^{2} - 12 a_{4} c_{0}^{2} \nu^{2} + 12 a_{4} c_{0} \nu - 3 a_{4}\right)}{8 \kappa \nu^{2}}
 \\\nonumber &\quad
    - \frac{3 \mobD c_{4} \left(4 a_{2} \nu^{2} - 12 a_{4} c_{0}^{2} \nu^{2}+ 12 a_{4} c_{0} \nu - 3 a_{4}\right)}{32 \kappa \nu^{2}}\\
    &\quad- \frac{\mobD \left(4 a_{2} \nu^{2} - 12 a_{4} c_{0}^{2} \nu^{2}+ 12 a_{4} c_{0} \nu - 3 a_{4}\right)^{2}}{64 \kappa \nu^{4}} - k \;.
    \label{eq:max_growth_rate}
    \end{align}
The spinodal is given by the concentration $c_0$ where the $\omega_\mathrm{max}$ becomes zero. We thus solve Eq. \ref{eq:max_growth_rate} for $c_0$ and find
\begin{align}
    c_\textrm{sp} 
    =	\frac{1}{2 \nu}
 \pm \sqrt{\frac{a_{2}}{3a_{4}} - \frac{2 \sqrt{\Lambda_\mathrm{d} k \kappa}}{3\Lambda_\mathrm{d} a_{4}}}
	 \;.
 \end{align}
\end{appendix}

\bibliography{lit_v2}%

\end{document}